\documentclass[twocolumn,showpacs,amsmath,amssymb,aps,prl,superscriptaddress]{revtex4-1}
\usepackage{xcolor,mathtools,graphicx,dcolumn,bm,hyperref}
\def\NEW#1{{\textcolor{black}{#1}}}

\newcommand{\be}{\begin{equation}}
\newcommand{\ee}{\end{equation}}
\def\ba{\begin{eqnarray}}
\def\ea{\end{eqnarray}}
\def\kk{{\bf k}}

\begin{document}
\preprint{1}

\title{Direct Evidence of a Dual Cascade in Gravitational Wave Turbulence}
\author{S\'ebastien Galtier}
\affiliation{Laboratoire de Physique des Plasmas, \'Ecole polytechnique, F-91128 Palaiseau Cedex, France}
\affiliation{Universit\'e Paris-Saclay, IPP, CNRS, Observatoire Paris-Meudon, France}
\affiliation{Institut universitaire de France}
\email{sebastien.galtier@universite-paris-saclay.fr}
\author{Sergey V. Nazarenko}
\affiliation{Institut de Physique de Nice, Universit\'e Nice-Sophia Antipolis, Parc Valrose, 06108 Nice, France}
\date{\today}

\begin{abstract}
We present the first direct numerical simulation of gravitational wave turbulence. General relativity equations are solved numerically in a periodic box with a diagonal metric tensor depending on two space coordinates only, $g_{ij} \equiv g_{ii}(x,y,t) \delta_{ij}$, and with an additional small-scale dissipative term. We limit ourselves to weak gravitational waves and to a freely decaying turbulence. We find that an initial metric excitation at intermediate wavenumber leads to a dual cascade of energy and wave action. When the direct energy cascade reaches the dissipative scales, a transition is observed in the temporal evolution of energy from a plateau to a power-law decay, while the inverse cascade front continues to propagate toward low wavenumbers. The wavenumber and frequency-wavenumber spectra  are found to be compatible with the theory of weak wave turbulence and the characteristic time-scale of the dual cascade is that expected for four-wave resonant interactions. The simulation reveals that an initially weak gravitational wave turbulence tends to become strong as the inverse cascade of wave action progresses with a selective amplification of the  fluctuations $g_{11}$ and $g_{22}$.
\end{abstract}

\maketitle

\paragraph*{Introduction.}
Wave turbulence (WT) is a state of a continuous medium with random mutually interacting waves of weak amplitude excited over a broad range of wavenumbers. The long-time statistical properties of such a medium have a natural asymptotic closure induced by the large separation of linear and nonlinear time scales \citep{Benney1966,Benney1967,Benney1967b}. The dynamics of WT is driven by kinetic equations which describe the redistribution of spectral densities via mainly three- or four-wave resonant interactions. The kinetic equations have two types of exact stationary power-law solutions: the zero-flux equilibrium thermodynamic spectra and the finite flux non-equilibrium Kolmogorov-Zakharov spectra \citep{ZLF92}. The latter solutions are much more interesting because they describe the spectral transfer of conserved quantities, such as energy or wave action, generally between a source and a sink \citep{Nazarenko11,NR11}. The direction of the cascade, direct or inverse, can be found by a numerical evaluation of the sign of the associated flux. The theory also offers the possibility to predict the Kolmogorov constant. All these properties makes WT a very interesting regime to understand in depth the mechanisms underlying turbulence. 

WT is of interest to many physical systems for which theoretical predictions have been made and verified numerically or experimentally. We have, among others, capillary waves \citep{BenneyN1967,Zakharov1967,Lommer2000,Pushkarev2000,Brazhnikov2001,Deike2014,Pan2014,Berhanu2019} and gravity waves \citep{Zakharov1966,Falcon2007,Clark2014,Aubourg2015} on fluid surfaces, inertial waves in rotating hydrodynamics \citep{Galtier2003,Bellet2006,Yarom2014,Clark2016,LeReun2017,Monsalve2020,Lereun2020,Yokoyama2021}, elastic waves on thin vibrating plates \citep{During2006,Boudaoud2008,Mordant2008,Cobelli2009,chibbaro2016,Hassaini2019}, optical waves in optical fibers \citep{Dyachenko1992,Laurie2012}, waves in Bose-Einstein condensate \citep{Nazarenko2006b,Proment2012}, Kelvin waves on quantum vortex filaments \citep{Kivotides2001,Kozik2004,Nazarenko2006}, magnetostrophic waves in geodynamo \citep{Galtier2014,Bell2019} and magnetohydrodynamic waves in space plasmas \citep{Galtier2000,Galtier2006,Meyrand2015,Meyrand2016,Meyrand2018,David2019}.
Recently, a theory of WT has been developed for gravitational waves (GW) \citep{GN2017}, a few years after their first direct detection \citep{Abbott2016}. 
\NEW{A promising application concerns the primordial universe shortly after the hypothetical initial singularity. During this period, GW can be produced by different mechanisms like e.g. first order phase transition \cite{Witten1984,Krauss1992} or the merger of primary black holes which can be formed from the primordial space-time fluctuations \citep{Carr2016}. A typical length scale of GW excitation can be $10^{-30}$m. Following this idea, a scenario of cosmological inflation has been proposed relying on presence of weak or strong GW turbulence and  rapid formation of a condensate via an inverse cascade \citep{Galtier2020b}. In this scenario, the initial strong GW bursts are quickly diluted as they propagate through the surrounding space, resulting in a statistically quasi-homogeneous GW field that is weakly or strongly non-linear depending on the strength and density of forcing events. For the weak WT case, a kinetic equation that describes the dynamics of energy and wave action via four-wave resonant interactions was derived. It has exact stationary scaling solutions for the one-dimensional (1D) isotropic spectrum of wave action: $k^{-1}$ for the direct energy cascade and  $k^{-2/3}$ for the inverse wave action cascade.}
Further, an explosive front propagation in the inverse cascade is predicted and observed numerically with a phenomenological nonlinear diffusion model where strongly local interactions are retained \citep{GNBT2019}. With this model, it is also shown that the non-stationary isotropic spectrum of wave action is slightly different from the Kolmogorov-Zakharov solution with a power-law index $\sim -0.6517$ instead of $-2/3$ for the stationary solution. However, this regime of WT is realized under assumptions that the initial condition consists of weak waves with random phases, and that the  nonlinearity remains weak  and the phases remain random during the evolution. Study of validity of such assumptions and robustness of the results with respect to more general (not necessarily wave dominated) initial conditions requires numerical simulations. Besides, the underlying nonlinear dynamics has never been explored in the physical space and this requires the use of numerical simulations. 

In this Letter, we report the first direct numerical simulations (DNS) of weak GW turbulence. This regime is studied in both physical and spectral spaces. 
Our study reveals that, despite having its own distinct properties, space-time turbulence behaves in a classical way from the point of view of the general turbulence theory.

\paragraph*{Einstein's equations.}

\begin{figure}
\includegraphics[width=1\linewidth]{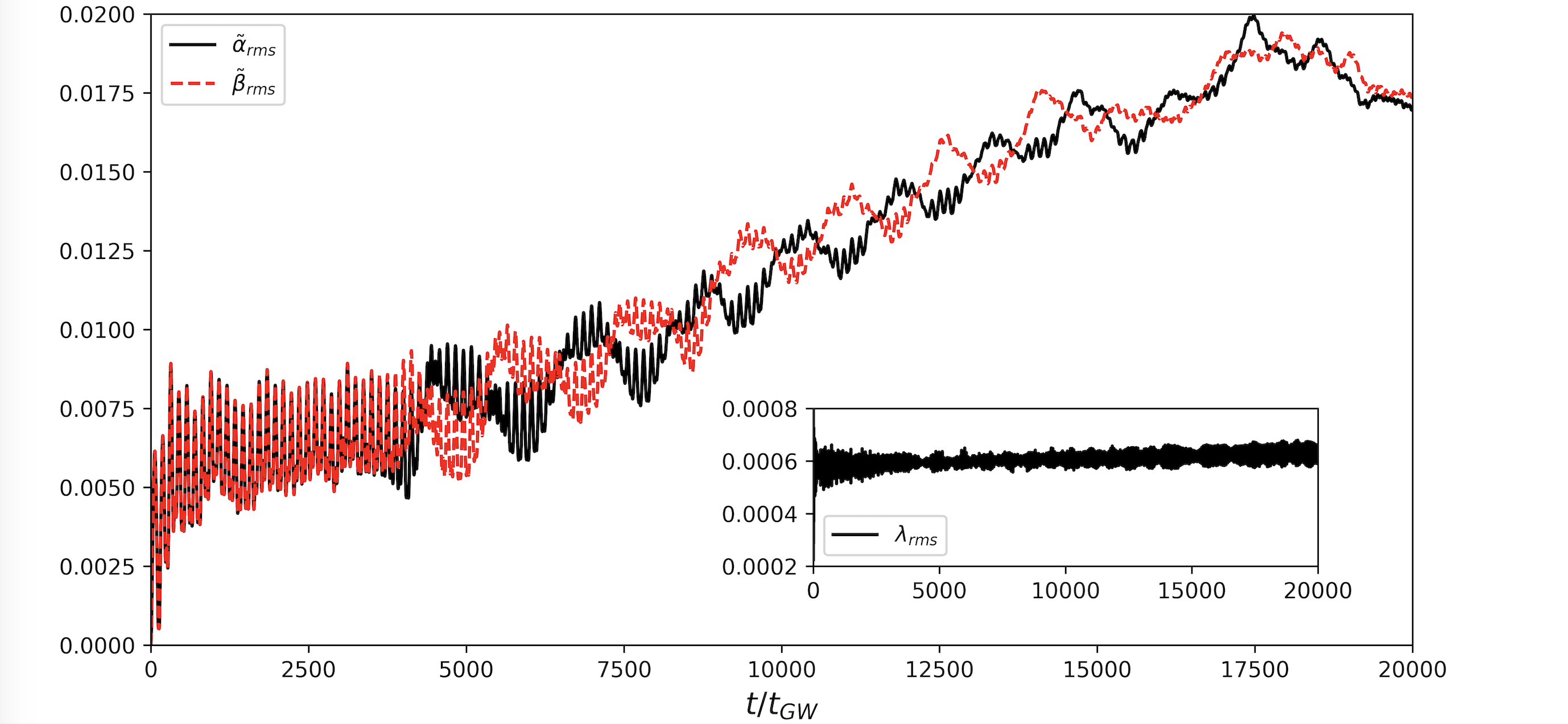}
\caption{Time evolution (rms values) of the basic fields $\tilde \alpha$, $\tilde \beta$ and $\lambda$ (insert).}
\label{fig1}
\end{figure}

Following \citep{GN2017,Hadad2014}, we shall consider the vacuum general relativity equations \citep{Einstein1915,Einstein1916} with the Hadad--Zakharov ($2.5+1$) diagonal metric: 
\ba
g_{00} &=& -(1+\tilde \gamma)^{2} e^{-2\lambda} , \quad g_{11} = (1+\tilde \beta)^{2} e^{-2\lambda} , \label{fields1} \\
g_{22} &=& (1+\tilde \alpha)^{2} e^{-2\lambda} , \quad g_{33} = e^{2\lambda} . \label{fields2}
\ea
The metric depends on the space-time variables $x$, $y$ and $t$ but is independent of $z$. 
\NEW{It can describe GW of any amplitude, however, we limit ourselves to weak GW amplitude, i.e.} 
$\tilde \alpha, \tilde \beta, \tilde \gamma, \lambda \ll 1$. Within this limit, Einstein's equations in the leading order are \citep{GN2017}:
\ba
\partial_x \dot{\tilde\alpha} = -2 \dot \lambda (\partial_x \lambda)  , &&  \quad \partial_y \dot{\tilde\beta} =  -2  \dot \lambda (\partial_y \lambda) ,  \label{eq1}\\
\partial_x \partial_y \tilde\gamma &=& -2 (\partial_x \lambda) (\partial_y \lambda) , \label{eq3} \\
\partial_t{ \left[( 1+\tilde\alpha+\tilde\beta- \tilde\gamma) \dot \lambda \right] } \, 
&=& \partial_x  \left[ ( 1+\tilde\alpha-\tilde\beta +\tilde\gamma) \partial_{x} \lambda \right]  \label{eq4} \\
&+& \partial_y  \left[ ( 1-\tilde\alpha+\tilde\beta + \tilde\gamma) \partial_{y} \lambda \right] , \nonumber
\ea
where we define $\dot f \equiv \partial_t f$. The linear solution of this system (see Eq. (\ref{eq4})) is a plus-polarized GW with a dispersion relation $\omega = k$ (\NEW{with the speed of light $c=1$ and $k = \vert \kk \vert$}) \citep{Maggiore08}. For the analysis  below, it is useful to recall the link between the canonical variable $a_{\kk}$ and the primary variables, $a_{\kk} = \sqrt{\frac{k}{2}} \lambda_{\kk} + i \sqrt{\frac{1}{2 k}} {\dot \lambda}_{\kk}$, where  $\lambda_{\kk}$ is the Fourier transform of $\lambda$ \citep{GN2017}.
The two-dimensional (2D) wave action spectrum \NEW{will be computed numerically using the relation $N(\kk) = \vert a_{\kk} \vert^{2}$ (homogeneity will be used as well as plane waves), and the 2D energy spectrum is $E(\kk) =  \omega N(\kk)$ \citep{GN2017}.}

\begin{figure}
\includegraphics[width=1\linewidth]{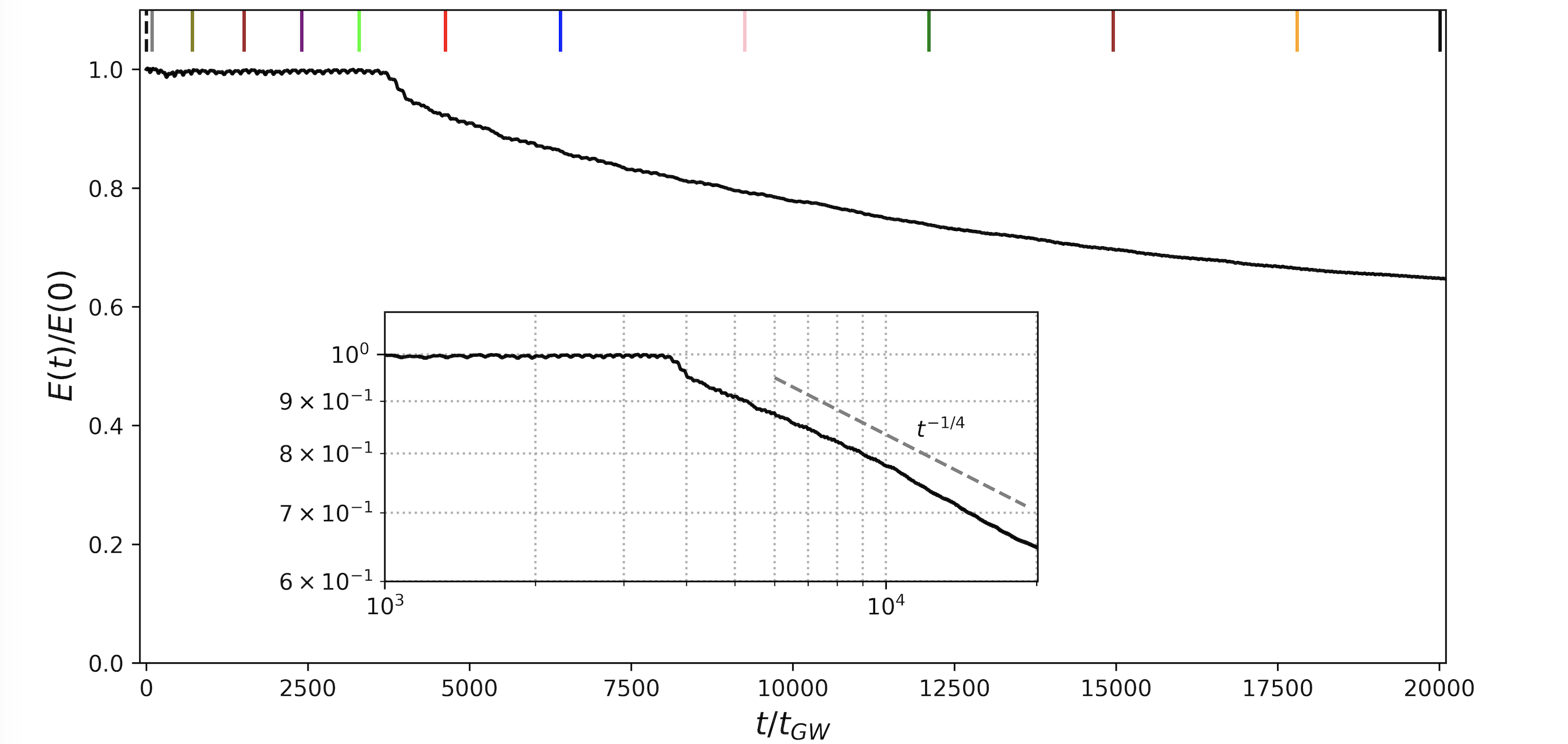}
\caption{Time evolution of the normalized energy. Insert: the same variation in log-log reveals a decay close to $t^{-1/4}$. The colored vertical lines at the top correspond to the times chosen to plot the spectra in Fig. \ref{fig3}.}
\label{fig2}
\end{figure}

DNS of Eqs. (\ref{eq1})--(\ref{eq4}) are performed with an additional  dissipative term acting at small scales to avoid numerical instabilities \citep{Pretorius2005}. This term, added in the RHS of Eq. (\ref{eq4}), takes the form $- \nu k^{4} {\dot \lambda}_{\kk}$ (for $k \ge k_{diss}$) in Fourier space \NEW{(see eg. \citep{During2006})}. 
\NEW{Physically, GW dissipation by matter is expected through e.g. Landau damping \citep{Baym2017}. (Note that the non-dissipative Einstein equations become invalid at the Planck length.) However, in our case, the dissipation has no precise origin and must be considered as a mechanism mimicking existence of a positive energy flux which would otherwise be blocked due to presence of the maximum wavenumber in our numerical method.}
We developed a pseudo-spectral code using FFTW3 routines, with periodic boundary conditions and dealiasing \citep{Shukla2019}. Since we deal with real fields only, the Fourier space is restricted to $k_y \ge 0$.  An Adams-Bashforth numerical scheme is used to integrate the nonlinear terms. For solving the double time derivative in Eq. (\ref{eq4}), an intermediate variable $\Lambda = \partial_{t} \lambda$ is introduced. We also introduce  intermediate variables $A = \partial_{t} \tilde \alpha$, $B = \partial_{t} \tilde \beta$ and $G = \partial_{t} \tilde \gamma$ for solving Eqs. (\ref{eq1}) and (\ref{eq4}) where several time-derivative terms appear. Therefore, in practice $8$ equations are numerically solved at each timestep $\Delta T$. The simulation shown is made with a spatial resolution of $512$ points in each direction, $\nu = 4 \times 10^{-11}$ and $k_{diss}=140$. Initially (at $t=0$) only $\lambda$ is excited around wavenumber $k_i=89$ \NEW{(in order to see the dual cascade)}, with random phases, and such that $\vert \lambda_{\kk} \vert^{2} = C (k^2 - 88^2)(90^2-k^2)$ with $k_{x,y} \in [88,90] / \sqrt 2$. The constant $C$ is chosen so that the total wave action is initially equal to $50$. The time will be normalized in the linear GW time unit $t_{GW} = 1 / \omega_{i} = 1/k_{i}$ (characteristic time of the initial excitation). 
\NEW{We take $\Delta T  = 10^{-5} t_{GW}$, which is relatively small but is necessary to avoid numerical instabilities and to ensure the conservation of invariants during the primary phase.}
The present simulation parameters appear to give the most representative illustration of the processes we study  (we have performed  a large number of simulations with various values of these parameters). 

\paragraph*{Results.}

Fig. \ref{fig1} shows the global evolution in time, i.e. the root mean square (rms) values, of the basic fields $\tilde \alpha$, $\tilde \beta$ and $\lambda$. We see that ${\tilde \alpha}_{rms}$ and ${\tilde \beta}_{rms}$ are strongly correlated with two distinct stages. At the first stage, until $\sim 3500 t_{GW}$, we see two almost identical  signals in which a slight growth is superimposed with oscillations  with a typical period of $100t_{GW}$. After that the signals become different from each other:  the high-frequency oscillations remain but in addition there appear long-scale oscillation (with a period about ten times longer) in which ${\tilde \alpha}_{rms}$ and ${\tilde \beta}_{rms}$ oscillate in counter-phase. At the large scale, both signals increase over time reflecting local amplifications of ${\tilde \alpha}(x,y)$ and ${\tilde \beta}(x,y)$. \NEW{Overall, $\tilde \alpha$ and $\tilde \beta$ behave like twin variables with similar dynamics.}
The behavior of ${\lambda}_{rms}$ (shown in the figure insert) and ${\tilde \gamma}_{rms}$ (not shown) is different with no significant increase in amplitude and faster oscillations with a period close to $t_{GW}$.

\begin{figure}
\includegraphics[width=\columnwidth]{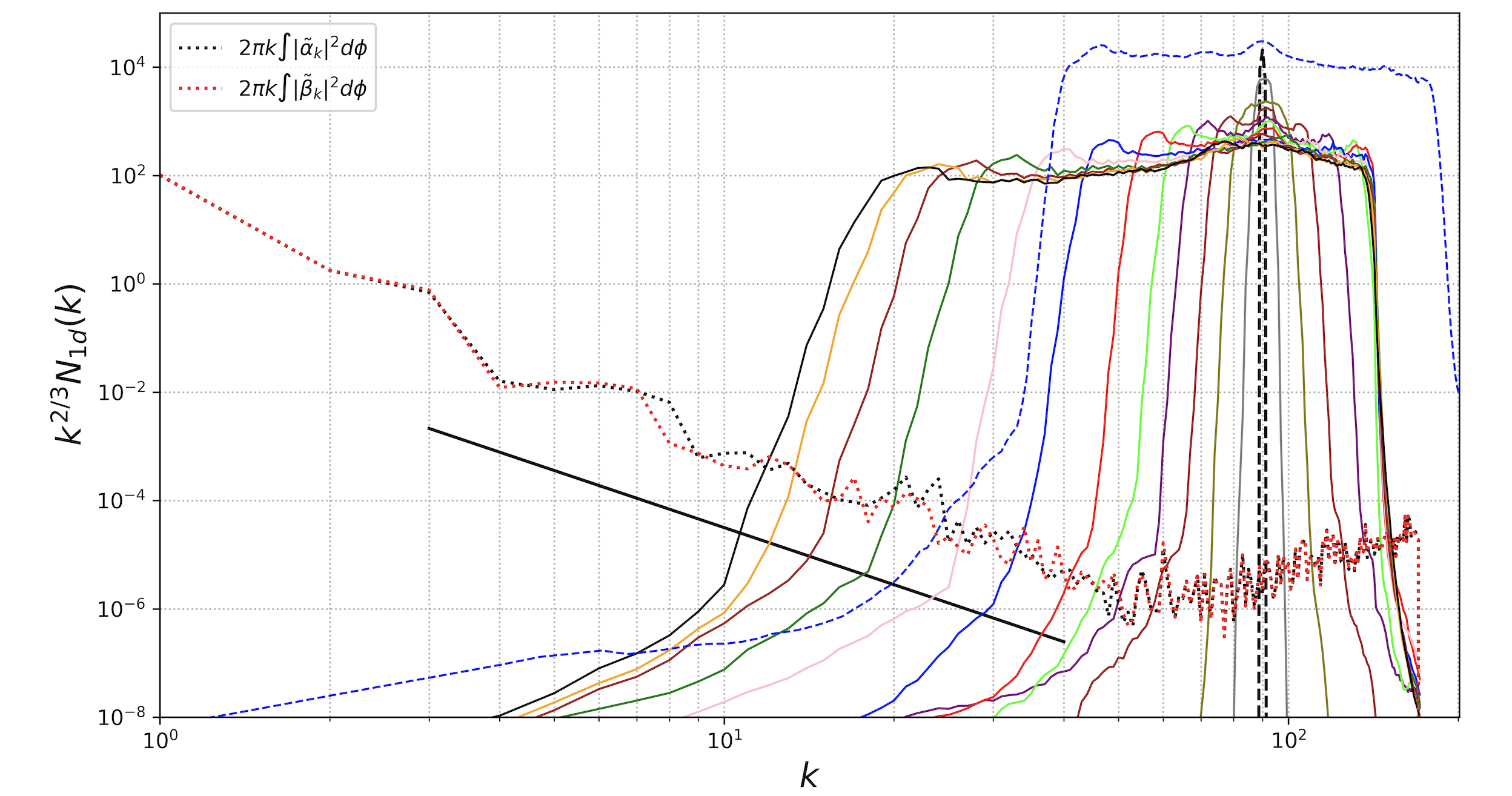}
\caption{Time evolution of $k^{2/3} N_{1d}(k)$ (solid lines) with the initial spectrum in black dashed line. Corresponding times are given in Fig. \ref{fig2}. The same spectrum, renormalized in scale and amplitude, is shown for an inviscid simulation at a resolution of $1024$ points (blue dashed line). The 1D spectra \NEW{$2\pi k \int_0^{2\pi} \vert \tilde \alpha_\kk \vert^2 d\phi$ and $2\pi k \int_0^{2\pi} \vert \tilde \beta_\kk \vert^2 d\phi$} are also shown (dotted lines) at the final time of the simulation (at a resolution of $512$ points) and compared with the power law $k^{-3.5}$.}
\label{fig3}
\end{figure}

In Fig. \ref{fig2}, we show the temporal evolution of the normalized energy $E$ (with initially $E(t=0)=28185$). Two phases are clearly present. First, the energy is conserved until $\sim 3500 t_{GW}$. This phase corresponds to the time interval needed to develop a direct energy cascade and to reach the small dissipative scales (see Fig. \ref{fig3}). This observation can be seen as an evidence of the accuracy of the numerical code (the wave action is also conserved during this phase). Then, the energy slowly decreases. To appreciate this decay law, a log-log plot is given in the insert: this reveals a power law decay in time. This type of behavior is classical in freely decaying (strong or weak) turbulence in presence of a direct cascade (see eg. \citep{Galtier1997,Bigot2008}). 

\begin{figure}
\includegraphics[width=1\linewidth]{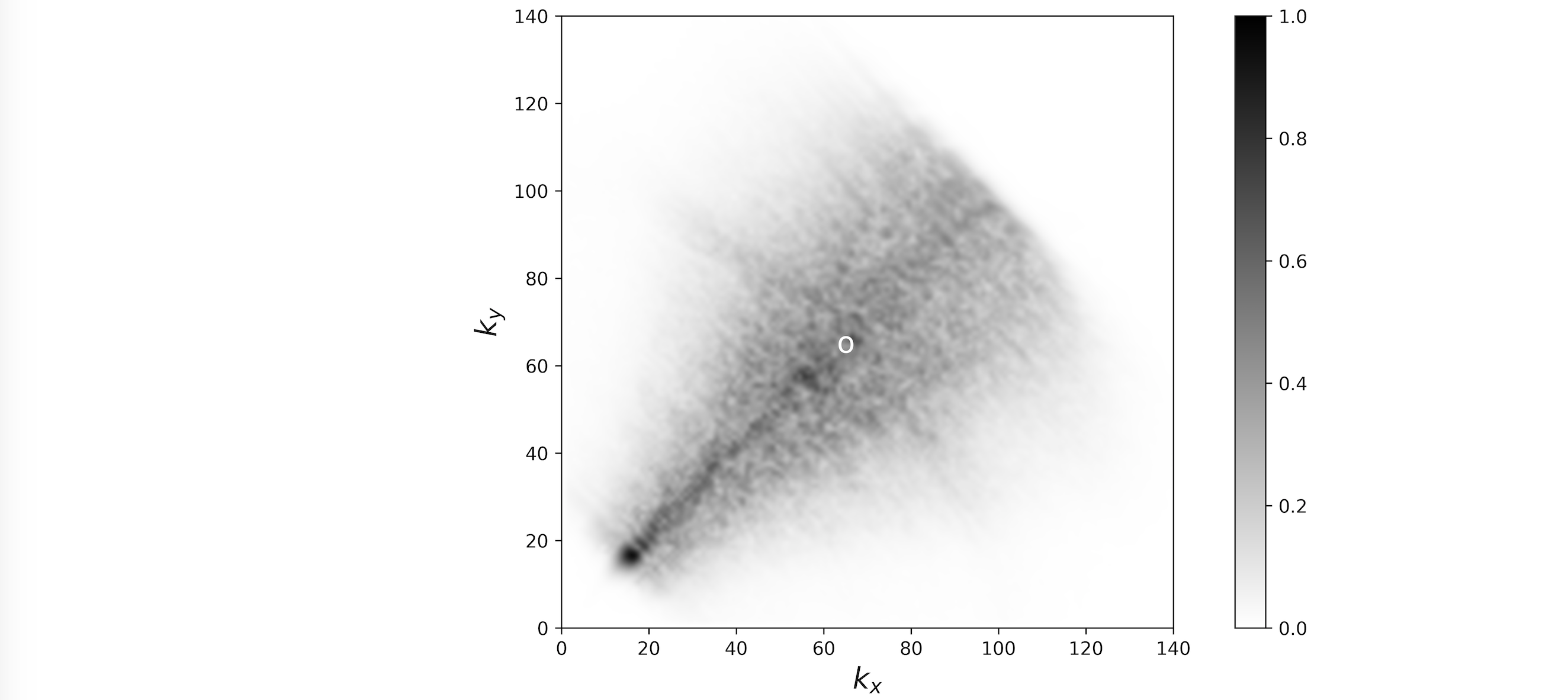}
\caption{$N(\kk)$ around the final time of the simulation (mean over 4 times). The center of the domain of initial excitation is indicated by the symbol 'o'.}
\label{fig4}
\end{figure}

The evolution in time of the 1D wave action spectrum \NEW{$N_{1d}(k)= 2\pi k \int_0^{2\pi} N({\bf k}) d\phi$, where $\phi$ is the polar angle in the k-space,}
is displayed in Fig. \ref{fig3}. The spectra are compensated by the theoretical prediction  $k^{-2/3}$ \citep{GN2017}. The propagation towards small scales is interpreted as a signature of a direct energy cascade. If we come back to Fig. \ref{fig2} we see that the  spectrum shown in light green is close to the end of the plateau: this is the moment when the dissipative scales are reached. Afterwards, the small-scale propagation of the spectrum is stopped. On the other hand, the development of the front is interpreted as an inverse cascade of wave action, which could be viewed as a strongly non-equilibrium Bose-Einstein condensation process. We see that beyond $3500 t_{GW}$ (from red curve onwards) the inverse cascade is preserved with a further expansion of the inertial range \NEW{characterized by a bump in the front propagation (see also Fig. \ref{fig4})}. We can already see the presence of a plateau and conclude that our result is qualitatively in agreement with the theoretical prediction. The 1D spectra \NEW{$2\pi k \int_0^{2\pi} \vert \tilde \alpha_\kk \vert^2 d\phi$ and $2\pi k \int_0^{2\pi} \vert \tilde \beta_\kk \vert^2 d\phi$} are shown in dotted lines: as expected, $\alpha$ and $\beta$ are much smaller than $\lambda$ \NEW{(since $N_{1d} (k) \sim k^2  \vert \lambda_\kk \vert^2 $)} in the inertial range of WT. 
 \NEW{However, we observe a significant selective amplification of $\alpha$ and $\beta$ at large scales which is not described by the weak WT theory.
Respectively, it means an amplification of the metric components $g_{11}$ and $g_{22}$ at the large scales while $g_{00}$ and $g_{33}$ remain in fast oscillations only (see also Fig. \ref{fig6}). For the GW, such large-scale variations of $g_{11}$ and $g_{22}$ are perceived as slow variation of the scale factor of the underlying space, i.e. its expansions and contractions.}
Superimposed to these plots, we also show the spectrum of an inviscid (i.e. $\nu=0$) simulation at $1024$ points resolution with an initial excitation at $k_i=128$ (with same timestep and type of initial condition). The simulation is stopped before reaching the smallest scales. The spectrum is renormalized in scale such that the initial excitation is moved to $89$. \NEW{The purpose of this plot is to confirm the dual cascade while we are still in the conservative phase, with a plateau at large scales and, at small scales, a wider inertial range compared to the $512^2$ simulation.}

In Fig. \ref{fig4} the normalized 2D wave action spectrum is shown around the final time of the simulation to appreciate the number of modes excited during these cascades. The presence of a bow at large wavenumbers can be seen as the signature of the dissipation which starts abruptly at  $k=140$. This plot reveals that a large domain in Fourier space is affected by the cascades which start around the symbol 'o'. Note that we are still far from the axes and no condensation is found. Note also a dark spot at the smallest excited wavenumbers corresponding to a spectrum \NEW{amplification}. It points out an overshoot nature of the propagating condensation front with a significant localized lump of wave action moving toward the smaller $k$'s. Also the 2D spectrum looks like a wedge/angle, which means that the propagation toward the large and the small scales takes place without significant spreading in the angular distribution of the wave propagation, i.e. without a full isotropization.

\begin{figure}
\includegraphics[width=\columnwidth]{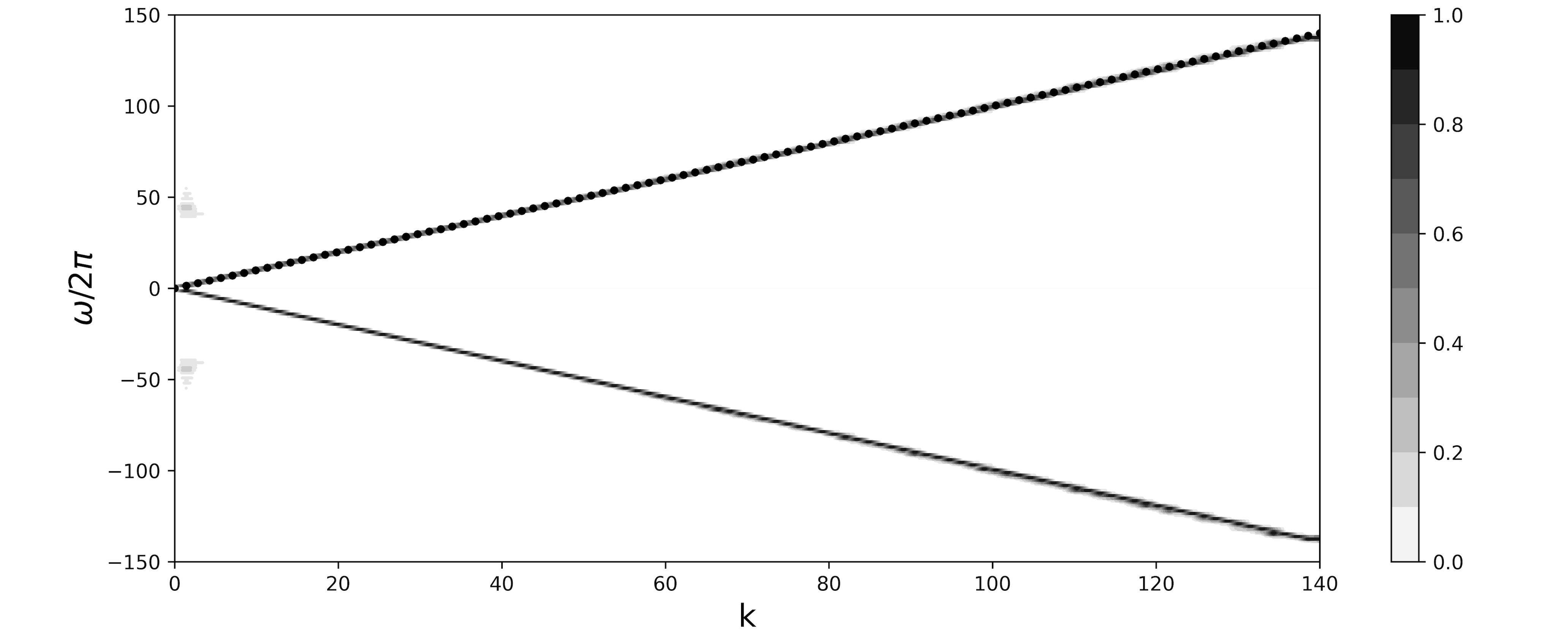}
\caption{$\omega$--$k$ spectrum of wave action. The dotted line corresponds to the dispersion relation of a GW.}
\label{fig5}
\end{figure}

\begin{figure}
\includegraphics[width=\columnwidth]{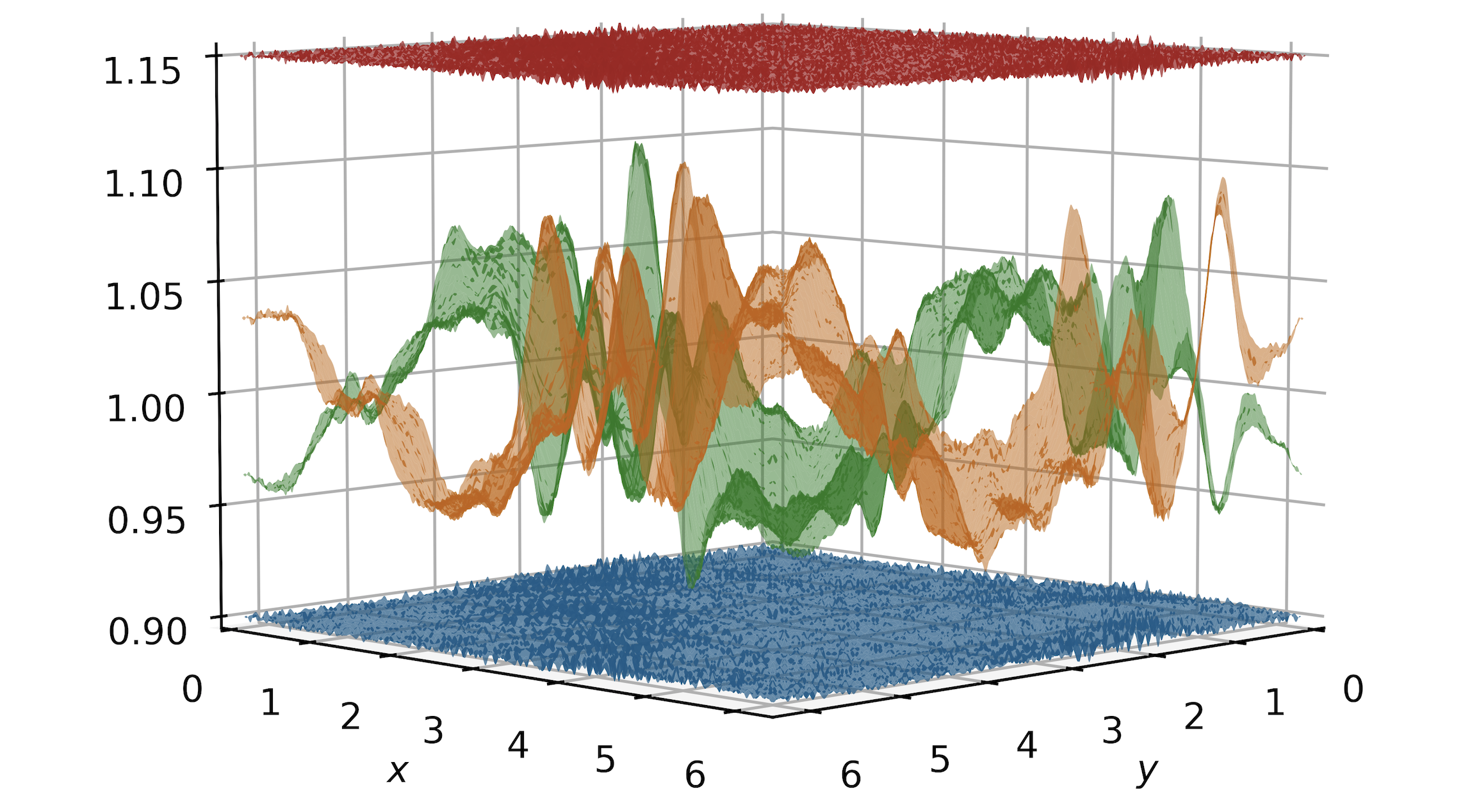}
\caption{Metric components $g_{00}$ (blue), $g_{11}$ (orange), $g_{22}$ (green) and $g_{33}$ (red) at the final time of the simulation. $g_{00}$ and $g_{33}$ have been vertically shifted  by $+1.9$ and $+0.15$, respectively.}
\label{fig6}
\end{figure}

WT is a state dominated by waves of weak amplitude; all non-wave initial disturbances eventually die out. Such a system is characterized by a very specific $\omega$--$k$ spectrum that concentrates near the dispersion relation curve of the wave in question \citep{Nazarenko2006,Cobelli2009,Clark2014,Yarom2014,Aubourg2015,Meyrand2015,LeReun2017,Bell2019,Hassaini2019}. It is also the case for GW turbulence as we can see in Fig. \ref{fig5} where a $\omega$--$k$ spectrum is plotted. This plot is obtained by taking the time evolution of the canonical variable $a_{\kk}$ (real part) for $t/t_{GW} \in [17000,20000]$. Then, we analyze signals corresponding to \NEW{$k \in [1,140]$, such that $k_x=k_y$.} A Fourier transform in time is then applied to each signal weighted with a Hamming function. The modulus squared of each signal normalized by its maximum is then plotted.  We see that a signal is obtained along the dispersion relation (dotted line) confirming the wave-like character of this turbulence. 


In Fig. \ref{fig6}, the four metric components are plotted at the final time of the simulation. We see that the dominant components are $g_{11}$ and $g_{22}$ whose large-scale oscillations are approximately in anti-phase. The two other metric components $g_{00}$ and $g_{33}$ are characterized by relatively small small-scale fluctuations around $-1$ and $+1$, respectively. These fluctuations remain relatively small during  the entire simulation. 
We have found that $g_{11}$ and $g_{22}$ behave similarly to each other during the simulation with mainly an anti-correlation and a gradual increase of the fluctuations. This observation is consistent with the behavior observed in Fig. \ref{fig1} where an increase of $\tilde \alpha$ and $\tilde \beta$ is also reported. The final time reported here is a reasonable limit to stop the simulation because the original equations used (\ref{eq1})--(\ref{eq4}) are only valid  for weak GW (which correspond to small fluctuations around $\pm 1$ of the metric components). Interestingly this final time, $t \sim 10^{4} t_{GW}$, corresponds to the expected time required to develop the weak turbulence regime when the small parameter $\epsilon$ used for the expansion is $\sim 0.1$: indeed, a phenomenological evaluation gives a typical cascade time $\tau_{\rm cascade} \sim \epsilon^{4} t_{GW}$ for four-wave interactions (whereas it would be $\sim \epsilon^{2} t_{GW}$ for three-wave interactions) \citep{Nazarenko11}. Note that this is an extremely slow time-scale from a numerical point of view that limits us to a relatively low spatial resolution (which however proved here to be sufficient to obtain physical results) {\footnote{{A run at resolution $1024^2$ takes about $6$ months.}}}. 

\paragraph*{Conclusion.}
In this Letter, we have reported the first DNS of GW turbulence. Specifically, the weak regime was studied for which analytical predictions exist. By using both physical and Fourier spaces we have been able to show that WT can emerge from an initial excitation of the space-time metric with a dual cascade of energy and wave action. This behavior is understood as the result of four-wave resonant interactions of the $2 \leftrightarrow 2$ type for which the wave action is an invariant. 
\NEW{Further, we have observed a new effect which is beyond the weak WT predictions--- emergence and continuous amplification of strong large-scale fluctuations of metric components $g_{11}$ and $g_{22}$ while the other components exhibit only weak small-scale oscillations. In particular the temporal component $g_{00}$ remains close to one which provides a natural cosmic time for weak GW turbulence.}
The amplification of  $g_{11}$ and $g_{22}$ limits our study since initially weak WT tends to become strong at large scales \citep{Galtier2020b}. In principle, the regime of strong WT can then be studied numerically with the metric (\ref{fields1})--(\ref{fields2}) \NEW{(whose form is preserved at all times as proved by \citep{Hadad2014}) by including all nonlinear terms of Einstein's equations.}

\NEW{The main conclusion of this work is that it is possible to produce turbulence in general relativity. Unlike the classical hydrodynamic turbulence, it does not consist of randomly interacting vortices but, rather, it takes a form of random interacting waves---the wave turbulence. Further we show that GW turbulence is a dual cascade system. Namely, in addition to the direct energy cascade,  there is an inverse cascade of wave action. The latter is important as it may shed light on the processes in early universe \citep{Galtier2020b}. We can also mention a strong similarity to elastic wave turbulence in the high tension limit. Indeed, both problems involve four-wave interactions with an inverse cascade of wave action \citep{Hassaini2019}. This similarity could be a motivation to pursue the comparison in an analog laboratory experiment to better understand strong GW turbulence in cosmology and the formation of a metric condensate.}


\bibliography{GWT}
\end{document}